\documentclass[aps,prl,twocolumn]{revtex4}
\usepackage{graphics}
\usepackage{graphicx}

\begin{document}


\title{Interaction renormalization and the plasma staircase}





\author{Alexander~V.~Milovanov${}^{1}$, Alexander~Iomin${}^{2}$, and Jens~Juul~Rasmussen${}^{3}$}

\affiliation{${}^1$ENEA National Laboratory, Centro~Ricerche~Frascati, I-00044 Frascati, Rome, Italy}
\affiliation{${}^2$Solid State Institute, Technion$-$Israel Institute of Technology, 32000 Haifa, Israel}
\affiliation{${}^3$Physics Department, Technical University of Denmark, DK-2800 Kgs.~Lyngby, Denmark}







\begin{abstract} We resolve an existing question concerning dynamical and structural stability of the ${\bf E} \times {\bf B}$ staircases in magnetically confined fusion plasma. The problem has stirred considerable interest in the literature due to a unique combination of its scientific and practical aspects. Here, we report a new approach to the problem by combining the paradigmatic Hasegawa-Wakatani model of electrostatic drift-wave turbulence with an idea of interaction renormalization, imported from quantum field theory. It is shown based on the Schr\"odinger-Newton equation with renormalized interaction constant that the plasma staircase is robust and stable as soon as the nonadiabaticity parameter, $\delta$, goes above a certain critical value, and is unstable otherwise. This critical value is an exact result of the interaction model and is found to be $\delta_c = \pi$. 
For $\delta < \delta_c$, the unstable system expands in the radial direction, resulting in a radial second moment that grows with time $t$ as $ \propto t^{2/3}$. The results obtained bear important implications with regard to the implementation of arrays of semipermeable transport barriers in tokamaks, suggesting a threshold condition for their generation. 
\end{abstract}


\maketitle

In a recent investigation of wave turbulence driven by inelastic wave scattering, Rosenhaus and Falkovich \cite{Falkovich} addressed the role of multiparticle and multimode correlations. They showed that the character of strong turbulence is determined by whether the effective four-wave interaction is enhanced or suppressed by collective effects. Consecutively they argued that the enhancement signals that strong turbulence is dominated by multiwave bound states, similar to confinement \cite{QED} in quantum chromodynamics.
In this Letter, we extend these ideas to magnetically confined fusion plasma and show that the effect of multiwave coherent structures (avalanches, blobs, holes) on the zonal flows in a tokamak can be characterized as a quantum chromodynamics-like antiscreening of the bare flow. 
A mean-field theory based on the Schr\"odinger-Newton equation (SNE) \cite{Penrose1,Penrose2,Penrose3,Bahrami} with renormalized interaction constant offers a concise theoretical understanding of a phenomenon often deemed too complex to be treated analytically: the plasma staircase \cite{DF2010,DF2015,DF2017,Horn2017}.     

A plasma staircase (or ${\bf E} \times {\bf B}$ staircase) is a quasiregular pattern of localized shear flows (``jets") and steepened pressure gradients (``steps") that form spontaneously in magnetized turbulent plasma, such as in tokamaks. It acts as a series of nested, semipermeable transport barriers \cite{DF2015,DF2017,Horn2017} that regulate turbulent transport via the velocity shear. The phenomenon is similar \cite{DF2010,Guillon} to the banded layering and alternating zonal jets observed in geophysical flows \cite{McIntyre}. Experimentally, the ${\bf E} \times {\bf B}$ staircases are identified for a large variety of near-marginal plasma conditions using ultrafast swept reflectometry \cite{DF2015,HL2A}. Theoretically, the plasma staircase exemplifies how the self-organization of plasma microturbulence in a tokamak breaks spatial and temporal symmetries through the formation of coherent structures \cite{JJR}, suggesting an exciting parallel \cite{PRE21,Caro} with quantum space-time crystals \cite{STC}.   

Despite progress in recent years \cite{Horn2017,Guillon,Gurcan,Korean,Malkov,Garbet,Nature}, the basic physics of staircase patterning is still largely debated in fusion community, with foundations from first principles lacking. The current interest in ${\bf E} \times {\bf B}$ staircases is amplified by the need to understand antidiffusion \cite{Acad}, a phenomenon where\textemdash contrary to standard diffusion\textemdash transport occurs up-gradient (from regions of low concentration to high concentration), signaling that a staircase system behaves as if it had negative viscosity or negative diffusion coefficient \cite{Hsu,Ashour}. An interesting new aspect here is an importance of nonlinear effects since in a good approximation the evolution of the plasma staircase can be described by the nonlinear Schr\"odinger equation (NLSE) with nonlocal order \cite{PRE21}. This interplay between nonlocality, nonlinearity and dispersion also occurs in other physical problems like dispersive soliton interactions \cite{Christ,Gaid,Lashkin}, energetic-particle physics \cite{Zonca06,Zonca15,Chen}, and many-body localization \cite{MBL,Abanin,Roy}, just to mention a few.  

Our model is as follows. In a poloidal cross-section, the staircase jets are represented as closed contours embracing the underlying magnetic flux surfaces (see Fig.~1). We consider each such contour of circular ${\bf E} \times {\bf B}$ motion as a nonlinear oscillator characterized by the radial position $x_j$, wave number $n_j = 2\pi / l_j$, and angular frequency $\omega_j = 2\pi / T_j$, where $T_j = l_j / u$ is the excursion period of a drift wave along a contour of length $l_j = \oint dy \sim 2\pi x_j$, $y$ is the coordinate along the contour, and $u = |{\bf E} \times {\bf B}|/{\bf B}^2$ is the drift velocity. In this representation, a system of coupled jets becomes a system of coupled nonlinear oscillators \cite{PRE21}. As such, it would obey the NLSE \cite{Comment}
\begin{equation}
i\hbar\frac{\partial\psi_n}{\partial t} = {\hat H}_{L}\psi_n + \Delta \hat H_{NL}\psi_n,
\label{DANSE} 
\end{equation} 
where $\hat H_{L}$ is the Hamiltonian of noninteracting oscillators \cite{Comm}, and $\Delta \hat H_{NL}$ incorporates eventual nonlinearities. In the above, $\psi_n = \psi (n,t)$ is a complex wave function, which characterizes the probability density to find an oscillator in a state with the wave number $n$ at time $t$, and the total probability is normalized to unity, i.e., $\sum_n |\psi_n|^2 = 1$. In writing the NLSE one assumes that the plasma staircase is a robust structure, which conserves energy on time scales long compared to the typical lifetime of plasma avalanches. It is assumed that the nonlinear interaction between the oscillators is provided by turbulent eddies that transfer energy and momentum across the interaction domain. The rate of this energy transfer is inversely proportional to the wave number difference, $|\Delta n| = n-n^\prime$ (because zonal flows are driven by the spatial divergence of the Reynolds stress, which acts as a momentum source or sink). Remembering that the energy density at point $n^\prime$ is directly proportional to $|\psi_{n^\prime}|^2$, one writes the nonlinear term as        
\begin{equation}
\Delta \hat H_{NL} \psi_n = \sum_{n^\prime}\beta\frac{|\psi_{n^\prime}|^{2}}{|n - n^\prime|}\psi_n,
\label{NLT} 
\end{equation} 
where $\beta > 0$ characterizes nonlinearity. Equation~(\ref{NLT}) mimics the interaction term in the SNE \cite{Penrose1,Penrose2,Penrose3,Bahrami}, with that modification that the $\beta$ value is taken to be positive (the SNE uses negative sign, due to the attractive character of gravitational interaction \cite{Segev}).    

\begin{figure}[t]
\includegraphics[width=0.5\textwidth]{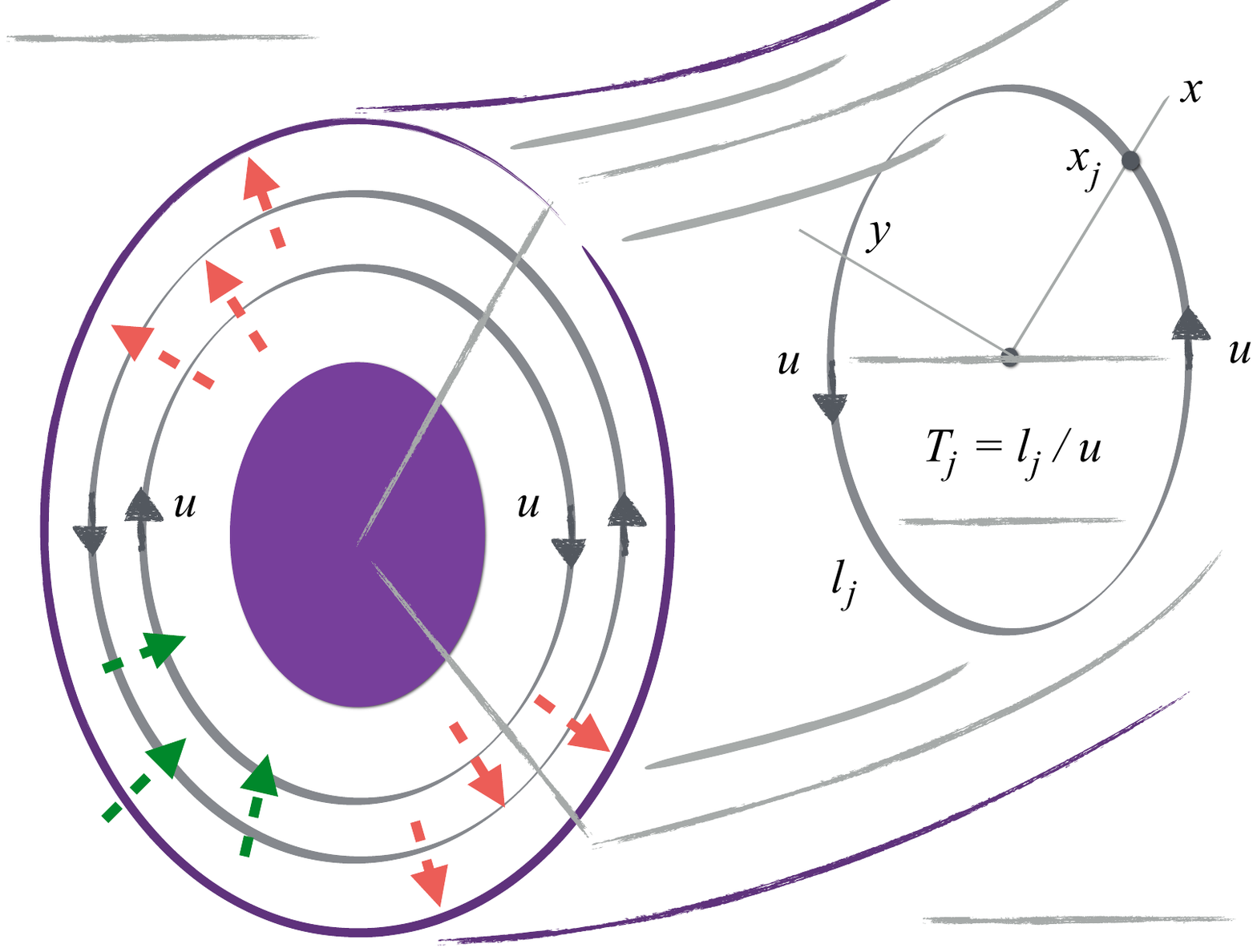}
\caption{\label{} The NLSE model. Elliptic contours (gray color) represent staircase jets circling around the magnetic flux surfaces. Each such contour is thought of as a nonlinear oscillator characterized by the radial position $x_j$, wave number $n_j = 2\pi / l_j$, and rotation period $T_j = l_j / u$, where $u$ is the ${\bf E} \times {\bf B}$ velocity. Dashed arrows represent respectively avalanches (red color) and voids (green color). Gray arrows indicate the direction of the ${\bf E} \times {\bf B}$  drift. The tokamak core is depicted as a central oval (violet color).      
}
\end{figure}

Our next step is to include the effect of coherent structures on the underpinning wave turbulence. A close inspection of numerical plots reported in Refs. \cite{PRE21,DF2015,DF2017,Horn2017} suggests that the ${\bf E} \times {\bf B}$ staircase is {naturally} surrounded by and interacts with a dynamic ``cloud" of meso-scale coherent structures: plasma avalanches\textemdash which emerge from the underlying turbulent background and propagate down-hill towards the plasma edge\textemdash and voids, or holes, which are coherent structures of reduced density and temperature, and which propagate up-hill from edge into core (see Fig.~1). Near a state of marginal stability, the newly born avalanches transport azimuthal momentum up the gradient of the azimuthal flow and drive the zonal-flow shear while moving outwards \cite{Xu}. During this process energy is transferred from the meso-scale coherent structures to the zonal flows via the turbulent Reynolds stress, resulting in nonlinear saturation of background turbulence, suppression of meso-scale fluctuations and intensification of the poloidal flow \cite{Long}. 

On the other hand, the increase in flow intensity implies that the corresponding pressure and density gradients have intensified. This feeds back the turbulence fluctuations, regenerating avalanches \cite{PLA14,Regen,PRE18}. The overall picture is that of the zonal flows continuously interacting with the underlying drift-wave microturbulence (and with each other), to which they are inseparably coupled \cite{Neglect}. We consider this ever pursuing, complex process of avalanche-jet zonal flow coupling as turbulent ``dressing" of the zonal flows by multiwave coherent structures (akin to quantum mechanical dressing of a quantum particle by virtual particles or quantum fluctuations). 

In the framework of the NLSE, it is a standard technique to account for the presence and dynamics of coherent structures by using (instead of the intrinsic coupling parameter, $\beta$) an effective (``renormalized") interaction parameter, i.e., 
\begin{equation}
\beta^\prime = \beta / [1 + \chi^\prime (\omega)],
\label{Ren} 
\end{equation} 
where $\chi^\prime (\omega)$ is the real part of the complex frequency-dependent susceptibility of the system, and $\beta$ is the intrinsic parameter [same as in Eq.~(\ref{NLT})]. Note that the effective interaction parameter in Eq.~(\ref{Ren}) is dispersive (its value depends on applied frequency) \cite{PRE25b}. 

With use of the Kramers-Kronig relation $\chi^\prime (\omega) = (1 / \pi) {P} \int \chi^{\prime\prime} (\omega^\prime)d\omega^\prime/(\omega^\prime - \omega)$, Eq.~(\ref{Ren}) implies that the renormalized values include the energy transfer processes (damping, turbulence drive) contained in the imaginary part ($\chi^{\prime\prime}$). $\chi^{\prime\prime}$ is obtained based on the known dispersion relation for the resistive drift wave \cite{JJR}
\begin{equation}
\omega (k_y, k) = \frac{k_y}{1+k^2} + i\delta\frac{k_y^2k^2}{(1+k^2)^3} - i\varepsilon\frac{k^4}{1+k^2},
\label{DR} 
\end{equation} 
where $k_y$ is the wavevector in poloidal direction (the $y$ direction in Fig.~1), $\delta$ is the nonadiabaticity parameter\textemdash which appears in the Hasegawa-Wakatani (HW) model \cite{HW1,HW2} of plasma edge turbulence and characterizes coupling between the potential and the density fluctuations,\textemdash and $\varepsilon$ incorporates eventual damping processes at the microscopic scales. We note in passing that the HW equations [and the associated dispersion relation~(\ref{DR}), which is obtained by linearizing these equations] serve as the standard paradigmatic fluid model for analyzing electrostatic resistive drift-wave turbulence in magnetically confined plasmas, despite the simplifications they carry \cite{JJR,Horton}. For $\delta\rightarrow 0$, the HW equations reduce to the Charney-Hasegawa-Mima equation \cite{HME}, which describes purely adiabatic drift-wave turbulence where particles follow potential surfaces, leading to strong zonal flow generation. On the contrary, when $\delta\rightarrow\infty$, the potential and density decouple, and the system behaves like 2D Navier-Stokes equations. 

In basic theory and simulations (e.g., Refs. \cite{Basu,PLA04}), one frequently uses $C$ instead of $\delta$. These two are related via $\delta = 1/C$, where $C= k^2_{\|}L^2_{\|}$ and contains via the scale length $L_{\|} = {(L_n T_e/m_e c_s \nu_{ei})}^{1/2}$ the parallel (along the $\bf B$ vector) resistivity. Here, $T_e$ and $m_e$ denote respectively the electron temperature and mass, $\nu_{ei}$ is the electron-ion collisional frequency, and $L_n = (\nabla w_0 / w_0)^{-1}$ is the length scale of the perpendicular background density gradient, where $w_0 = w_0 (x)$ is the background density. 

Equation~(\ref{DR}) shows that the deviation from adiabaticity, given by the parameter $\delta$, leads to an instability with a maximum growth rate $\gamma_{\max} \approx \delta / 8$ at $k\approx 1$. It is understood that $\gamma_{\max}$ induces a peak in the imaginary part of the complex susceptibility function, where the magnitude of $\chi^{\prime\prime}$ is maximized. We denote the peak value by $\chi_{\max}^{\prime\prime}$. Based on the dispersion relation in Eq.~(\ref{DR}) it is straightforward to demonstrate that $\chi_{\max}^{\prime\prime} \approx -\delta/2$, where the minus sign indicates that the energy is pumped into the flow via the action of instability. The peak value is typically half of the static susceptibility, i.e., $\chi_{\max}^{\prime\prime} \approx \chi_0 /2$ \cite{Static}, yielding $\chi_0 \approx -\delta$. This is to be expected as $\delta$ characterizes (together with $\chi_0$) the responsiveness of the density fluctuations to low-frequency perturbations in the electrostatic potential. In the Kramers-Kronig integral, one identifies $\chi^{\prime\prime} (\omega^\prime)$ with the peak value, i.e., $\chi^{\prime\prime} (\omega^\prime) \approx \chi_{\max}^{\prime\prime}$, making it possible to express the real part as $\chi^\prime (\omega) = -\mu \int d\omega^\prime/(\omega^\prime - \omega)$, with $\mu = \delta/2\pi$. The integration over $d\omega^\prime$ is performed in the limits from $\sim \omega$ to $\Delta \omega$, where $\omega = 2\pi u / l$ is the characteristic drift-wave excursion frequency in the poloidal plane, $u$ is the ${\bf E} \times {\bf B}$ velocity, $l$ is the characteristic poloidal span of the drift-orbital motion, and $\Delta\omega$ is the frequency spread of the staircase. Thus we have, with logarithmic accuracy, $\chi^\prime (\omega) \simeq -\mu \log |\Delta \omega / \omega|$. Translating frequencies into wave numbers using $\omega = nu$, one gets $\chi^\prime (n) \simeq -\mu \log |\Delta n / n|$, where $\Delta n$ designates the range of staircase self-organization in wave number space. If, however, the frequency spread $\Delta\omega$ is large compared to $\omega$, i.e., $\Delta\omega\gg\omega$, then $\Delta n$ will be respectively large compared to $n$, i.e., $\Delta n \gg n$, permitting one to express the real part of $\chi$ as $\chi^\prime (\Delta n) \simeq -\mu \log |\Delta n|$, without keeping an eye on the specific $n$ value. Writing, under the logarithm sign, the spatial spread $\Delta n$ as $|n-n^\prime|$, where $|n-n^\prime| \gg 1$, one obtains $\chi^\prime (n, n^\prime) \simeq -\mu \log |n-n^\prime|$, from which, with the aid of Eq.~(\ref{Ren}),   
\begin{equation}
\beta^\prime = \beta / (1 -\mu \log |n-n^\prime|).
\label{Ren2} 
\end{equation} 

One sees that the renormalized $\beta$ takes the form of effective interaction parameter in quantum chromodynamics, where it accounts for antiscreening and asymptotic freedom \cite{Gross,Politzer}. 

This correspondence with quantum chromodynamics is not really surprising. It elucidates the enhancing effect of meso-scale fluctuations on the zonal flows, showing that the interaction strength grows with distance. The hint is that the interactions are mediated by nonlocal structures (avalanches), rather than point-wise particles. This antiscreening process is controlled self-consistently by the nonadiabaticity parameter $\delta = 1/k^2_{\|}L^2_{\|}$ and finds the energy reservoir in the surrounding turbulent background. 

Replacing $\beta$ by $\beta^\prime$ in Eq.~(\ref{NLT}), one gets 
\begin{equation}
\Delta \hat H^\prime_{NL} \psi_n = \sum_{n^\prime} \frac{\beta}{1 -\mu \log |n-n^\prime|} \frac{|\psi_{n^\prime}|^{2}}{|n - n^\prime|}\psi_n,
\label{NLT-R} 
\end{equation} 
which incorporates the avalanche-jet zonal flow coupling (and the associated negative-viscosity increment) via a logarithmic correction to $\beta$. If $\mu \ll 1$, i.e., the system is in the HW regime, not Navier-Stockes regime, then one can substitute, in the first order of Taylor expansion, the logarithmic form $1 -\mu \log |n-n^\prime|$ with the algebraic form $|n-n^\prime|^{-\mu}$, i.e., $1 -\mu \log |n-n^\prime| \approx |n-n^\prime|^{-\mu}$, yielding
\begin{equation}
\Delta \hat H^\prime_{NL} \psi_n \simeq \beta \sum_{n^\prime} \frac{|\psi_{n^\prime}|^{2}}{|n - n^\prime|^{1-\mu}}\psi_n.
\label{NLT-RR} 
\end{equation} 
One sees that the effect of coherent structures is equivalent to some modification of the kernel function, which acquires a sublinear shape, i.e., $\propto 1/ |n - n^\prime|^{s}$, with the power exponent $s = 1-\mu$, where $\mu= \delta/2\pi \ll 1$. If one considers the sublinear kernel $\propto 1/ |n - n^\prime|^{s}$ as a ``slow" function, one can replace $|n - n^\prime|$ with the characteristic spread $\Delta n$ and thus take $1 / |n-n^\prime|^{1-\mu} \simeq 1/|\Delta n|^{1-\mu}$ out of the summation sign in Eq.~(\ref{NLT-RR}). Then one is left with the sum $\sum_{n^\prime} |\psi_{n^\prime}|^{2}$, which is immediately seen to converge to 1 by virtue of the conservation of the total probability, i.e., $\sum_{n^\prime} |\psi_{n^\prime}|^{2} = 1$. On the other hand, if the field is spread over $\Delta n \gg 1$ states, then the same conservation of the total probability would imply that the field is small, i.e., $|\psi_{n}|^{2} \sim 1/\Delta n$, from which $1/|\Delta n|^{1-\mu} \simeq |\psi_n|^{2(1-\mu)}$. Using $s = 1-\mu < 1$, one gets 
\begin{equation}
\Delta \hat H^\prime_{NL} \psi_n \simeq \beta|\psi_n|^{2s}\psi_n.
\label{NLT-RF} 
\end{equation} 
Equation~(\ref{NLT-RF}) is a mean-field reduction of the interaction Hamiltonian in Eq.~(\ref{NLT-R}). It includes the antiscreening (coupling to avalanches) processes via the deviation of the $s$ value from 1 in $|\psi_n|^{2s} \equiv (|\psi_n|^{2})^s$. Combining Eqs.~(\ref{DANSE}) and~(\ref{NLT-RF}), one is led to the NLSE with subquadratic nonlinearity \cite{PRE19,EPL23,PRE23}
\begin{equation}
i\hbar\frac{\partial\psi_n}{\partial t} = {\hat H}_{L}\psi_n + \beta |\psi_n|^{2s} \psi_n. 
\label{SNLSE} 
\end{equation} 
In the context of staircase physics, the NLSE~(\ref{SNLSE}) was proposed (without derivation) in Ref. \cite{PRE21}, yet the meaning of the subquadratic power was not obtained then. Here, we find that the subquadratic power is the direct mathematical consequence of the processes of antiscreening taking place. It represents in a reduced, compact form the integral effect of avalanche-jet zonal flow coupling, by relating the coupling strength $(\mu)$ to the $s$ value via $s = 1-\mu$, where $\mu = \delta/2\pi \ll 1$. In the adiabatic limit $\delta \rightarrow 0$, one gets $\mu \rightarrow 0$ and $s\rightarrow 1$, for which the standard NLSE is inferred. 

By mapping the nonlinear term onto a ``forest" of Cayley trees one shows \cite{EPL23,PRE23} that the subquadratic nonlinearity acts as a system of communication links (percolation-like) between the eigenstates involved. For $s < 1/2$, the global communication pattern is disconnected and therefore inadequate to induce global behavioral change. On the contrary, if $s > 1/2$, then the communication pattern is topologically connected, with outstanding dynamical implications \cite{EPL23,PRE19}. As a result, the nonlinear field with long-range communications between the eigenstates expands into the ambient space in accordance with the subdiffusive law \cite{PRE23} 
\begin{equation}
\langle|\Delta n|^2 (t)\rangle \propto t^{1/(1+s)}, \ \ \ t\rightarrow+\infty.
\label{SP} 
\end{equation} 
Thus it follows that the NLSE~(\ref{SNLSE}) contains a localization-delocalization transition, which occurs at $s=1/2$ exactly. If $s=1$, then Eq.~(\ref{SP}) reduces to $\langle|\Delta n|^2 (t)\rangle \propto t^{1/2}$, which is frequently found in systems with chaotic subdiffusion \cite{Many,QW,PRE17}. Translating $s$ into the $\delta$ values using $s = 1-\delta/2\pi$, one sees that there is a critical (threshold) value of the nonadiabaticity parameter, $\delta_c$, for which the nonlinear field in Eq.~(\ref{SNLSE}) transits from a localized ($\delta > \delta_c$) into delocalized ($\delta < \delta_c$) state. This critical value is immediately seen to be given by $\delta_c = \pi$. 

One concludes that the nonadiabaticity parameter must be greater than $\delta_c$ in order to hold the ${\bf E} \times {\bf B}$ staircase in a spatially bound state. 
We associate this critical value of $\delta$ with the antiscreening effect of plasma avalanches, which suppresses (for $\delta > \delta_c$) the kinetic instabilities driving chaotic spreading.  

In a basic physics of magnetically confined fusion plasma it is shown \cite{Zonal,Zonal06} that zonal flows are driven {\it exclusively} by nonlinear interactions, which transfer energy from the finite-$n$ drift waves to the $n=0$ flow. Usually, such nonlinear interactions are three-wave triad couplings between two high $\bf k$ running drift waves and one low $n$ zonal flow excitation \cite{Zonal}. The process leads universally to a subdiffusive spreading of the zero-frequency waves in wave number space in accordance with \cite{PRE24,PRE25a}
\begin{equation}
\langle|\Delta n|^2 (t)\rangle \propto t^{2/3}, \ \ \ t\rightarrow+\infty.
\label{SPE} 
\end{equation}
Comparing Eqs.~(\ref{SP}) and~(\ref{SPE}), one demands $1/(1+s) = 2/3$, from which $s=1/2$. Thus it follows that the staircase system occurs {\it exactly} at the delocalization border between localized and delocalized states. It fluctuates around this border by emitting and reabsorbing plasma avalanches and voids, which regulate the intensity of poloidal flows via the turbulent Reynolds stress [mathematically, via the dependence of the parameter $\delta$ on the perpendicular density gradient, i.e., $\delta \propto L_n^{-1} = \nabla w_0 (x) / w_0 (x)$]. This coupling between avalanches and flows acts as a feedback mechanism that ensures that the point of localization-delocalization transition is robust and self-organized. In complex dynamics, this tendency of interaction-dominated, driven dissipative systems to evolve into a critical state is known as self-organized criticality (SOC). The notion was introduced by Bak, Tang, and Wiesenfeld (BTW) in their seminal work \cite{BTW} and received an outstanding interest in the literature since. Here, we propose based on Eqs.~(\ref{SP}) and~(\ref{SPE}) that the plasma staircase operates as a SOC system in vicinity of the delocalization point $s=1/2$, $\delta = \delta_c$. We hasten to note that the power-law reduced distributions of plasma avalanches over their sizes\textemdash obtained in Refs. \cite{PRE21,PRE25a}\textemdash differ appreciably from those characterizing the BTW sand-pile \cite{BTW} and its modifications \cite{CSF}, suggesting a different universality class.  

{In summary}, we have shown that multiwave coherent structures (avalanches, blobs, holes) produce a quantum chromodynamics-like antiscreening to azimuthal flows in a tokamak. This antiscreening process is the direct consequence of up-gradient propagation of plasma avalanches and mathematically corresponds to a negative-viscosity increment in the nonlinearity parameter. A simple (SNE based) model for the coupled avalanche-jet zonal flow system delivers a concise theoretical understanding of the plasma staircase in good agreement with experimental results. The model predicts that there exists a critical value of the nonadiabaticity parameter $\delta = 1/C$, such that above that value the staircase system is stable and spatially localized, while below which it undergoes a phase transition from a localized into a delocalized state. This critical value is found to be $\delta_c = \pi$ (within the framework of the HW model). Also we have shown that the plasma staircase occurs exactly at the edge of its delocalization transition, where it is marginally stable against disturbances (via the dependence of $\delta$ on the perpendicular density gradient)\textemdash signaling a fundamental connection to SOC. 
We expect this behavior to extend beyond the paradigmatic HW model to more realistic turbulence models, with a similar critical parameter $\delta_c$. 

More generally and more importantly, our results validate the use of arrays of semipermeable transport barriers (i.e., the plasma staircase \cite{DF2010,DF2015}) as a prospective and efficient means to confine hot thermonuclear plasma in a tokamak. An essential  key element here is an understanding that the plasma staircase implies a nontrivial stability range ($\delta > \delta_c$), and that structural stability is provided dynamically via coupling to propagating coherent structures, avalanches and voids. The phenomenon is captured universally by the NLSE with subquadratic nonlinearity, where the exponent of subquadratic power incorporates the processes of antiscreening taking place.       

\acknowledgments
A.V.M. thanks G. Dif-Pradalier and participants of the 2025 Festival de Th\'eorie for insightful discussions on the staircase problem. 
Also A.V.M. thanks the Isaac Newton Institute for Mathematical Sciences, Cambridge, U.K., for support and hospitality during the program ``Anti-diffusive dynamics: from sub-cellular to astrophysical scales." This work was supported by EPSRC grant No. EP/R014604/1. Partial support was received from a grant from the Simons Foundation.


\begin{thebibliography}{}
\bibitem{Comm}
A suitable form of $\hat H_{L}$ is reported in Ref. \cite{PRE21}. It is tacitly assumed that $\hat H_{L}$ possesses a full basis of exponentially localized eigenstates and that the eigenspectrum is discrete and dense. This is ensured by the fact that the drift waves in a tokamak are attracted to rational flux surfaces, inducing a form of radial localization (akin to Anderson localization in disordered lattices).
\bibitem{Neglect}
Here we neglect microscopic dissipation and losses to boundaries. Near a state of marginal stability, these both a compensated by the energy source in tokamak core.
\bibitem{Static}
The relationship between the static susceptibility and the peak value ($\chi^{\prime\prime}_{\max}$) of the imaginary part is most directly defined by the Debye relaxation model. In this model, the complex susceptibility is written as $\chi (\omega) = \chi_0/(1-i\omega\tau)$, where $\chi_0$ is the static susceptibility. The imaginary part $\chi^{\prime\prime} (\omega) = \chi_0\omega\tau/[1+(\omega\tau)^2]$ peaks when $\omega\tau =1$, where $\tau$ is the relaxation time, leading to $\chi^{\prime\prime}_{\max} = \chi_0 / 2$.
\end{thebibliography}


\end{document}